\documentclass[aps,pre,twocolumn]{revtex4} 
\usepackage{amsmath, epsfig} 
 
\newcommand{\vp}{{\bf{p}}} 
\renewcommand{\vr}{{\bf{r}}} 
\newcommand{\vk}{{\bf{k}}} 
\newcommand{\et}{\eta} 
\newcommand{\ze}{\zeta} 
\newcommand{\ps}{\psi} 
\newcommand{\la}{{\lambda \!\!\!\!\, {}^-}} 
\newcommand{\trho}{\tilde{\rho}} 
\newcommand{\real}{\: {\mathcal R}\!e \!} 
\newcommand{\imag}{\: {\mathcal I}\!m \!} 
 
\newcommand{\T}{\mathrm{T}} 
\newcommand{\vet}{\mbox{\boldmath$\et$}} 
\newcommand{\vOmega}{\mbox{\boldmath$\Omega$}} 
\newcommand{\unit}{{\bf 1}} 
\newcommand{\vP}{{\bf P}} 
\newcommand{\vQ}{{\bf Q}} 
\newcommand{\vrho}{\mbox{\boldmath$\rho$}} 
\newcommand{\vtrho}{\mbox{\boldmath$\trho$}} 
\newcommand{\vSig}{\mbox{\boldmath$\Sigma$}}

\begin{document} 
 
\title{Lyapunov Spreading of Semi-classical Wave Packets for the Lorentz 
  Gas: \\ Theory and Applications} 
 
\author{Arseni Goussev and J. R. Dorfman} 
 
\affiliation{Institute for Physical Science and Technology and Department 
  of Physics \\ University of Maryland \\ College Park, MD 20742 USA} 
 
\date{\today} 
 
\begin{abstract} 
We consider the quantum mechanical propagator for a particle moving in a 
$d$-dimensional Lorentz gas, with fixed, hard sphere scatterers. To 
evaluate this propagator in the semi-classical region, and for times less 
than the Ehrenfest time, we express its effect on an initial Gaussian wave 
packet in terms of quantities analogous to those used to describe the 
exponential separation of trajectories in the classical version of this 
system. This result relates the spread of the wave packet to the rate of 
separation of classical trajectories, characterized by positive Lyapunov 
exponents. We consider applications of these results, first to illustrate 
the behavior of the wave-packet auto-correlation functions for wave packets 
on periodic orbits. The auto-correlation function can be related to the 
fidelity, or Loschmidt echo, for the special case that the perturbation is 
a small change in the mass of the particle. An exact expression for the 
fidelity, appropriate for this perturbation, leads to an analytical result 
valid over very long time intervals, inversely proportional to the size of 
the mass perturbation. For such perturbations, we then calculate the 
long-time echo for semi-classical wave packets on periodic orbits.
% ------- To be removed for PRE -------
This paper also corrects an earlier calculation for a quantum echo, included
in a previous paper, (nlin.CD/0307025). We explain the reasons for this correction.
% -------------------------------------  
\end{abstract} 
 
\maketitle

\section{Introduction} 
The search for signatures of classical chaos in corresponding quantum 
systems is one of the main themes of quantum chaos. The literature in this 
subject is large, much of it described in recent books by Haake and by 
St\"ockmann \cite{haake,stock}. Here we consider a simple version of this 
question, namely, we describe a semi-classical calculation of the short 
time spreading of a wave-packet for the quantum version of a classically 
chaotic system and show, as one might expect, that it is governed to a 
large extent by the Lyapunov exponents characterizing the exponential 
separation of close trajectories of the classical system. We consider a 
Lorentz gas, which consists of a particle, or a collection of 
non-interacting particles, moving in $d$-dimensions among a collection 
of fixed scatterers, generally taken to be $d$-dimensional hard spheres, 
with finite horizon. The case where the scatterers are centered at the 
vertices of a regular lattice is the Sinai billiard. Related work on this 
problem has been done by Wirzba \cite{wirz}, which will be mentioned in 
Section II.  
 
We consider the propagator for a semi-classical particle moving among the 
scatterers. The de Broglie wave length of the moving particle is taken to 
be small compared both to the size of a scatterer and to the average 
distance between scatterers. The propagator is evaluated by semi-classical 
methods for time intervals where a number of collisions take place. As long 
as the wave-packet remains small, its spreading with time is governed by a 
set of equations that are the quantum counterparts to the curvature 
equations of Sinai and co-workers \cite{sinai} that determine the Lyapunov 
exponents and Kolmogorov-Sinai (KS) entropy for the classical system. We 
can then easily relate the spreading of the wave-packet to the classical 
Lyapunov exponents.  
 
Next we apply this result to a calculation of the wave-packet 
auto-correlation function (the return probability), $C(t)$, defined by 
\begin{equation} 
C(t)=\left| \langle 0|e^{-itH/\hbar}|0 \rangle \right|^2, 
\label{1.1} 
\end{equation}
where $H$ is the Hamiltonian of the system and $|0\rangle $ is an initial 
quantum state. We find that, for wave packets on periodic orbits, this 
function exhibits a series of maxima, with amplitudes decreasing mainly 
exponentially with time, as $\exp(-h_\mathrm{KS} t)$, where 
$h_\mathrm{KS}=\sum_i \lambda_i$ is the classical KS-entropy, and the 
$\lambda_i$ are the positive Lyapunov exponents for the corresponding 
periodic orbits. This type of decay of the auto-correlation function was 
first described by Heller \cite{heller}. Here we also calculate the 
coefficient in front of the exponential. This coefficient has a 
sub-exponential time dependence. The auto-correlation function exhibits a 
set of maxima separated by deep minima that appear for simple physical 
reasons, as we explain in Section III. 
 
These results are limited in applicability to times {\it less than the 
Ehrenfest time}, which is the time necessary for a wave packet to spread to 
the size of a scatterer. However, there is an application of them to the 
Loschmidt echo, or quantum fidelity, of a special type which is valid for a 
much larger time interval, greater than the Ehrenfest time. The Loschmidt 
echo \cite{peres,past,cucc}, $M(t)$, is defined by 
\begin{equation}
M(t) =\left| \langle 0|e^{it(H+\Sigma)/\hbar}e^{-itH/\hbar}|0 \rangle 
\right|^2. 
\label{1.2} 
\end{equation}
Here $H$ is the Hamiltonian for the system, $\Sigma$ is a small perturbing 
Hamiltonian, and $|0 \rangle$ is some initial quantum state. For the case 
that $H$ is the Lorentz gas Hamiltonian with hard sphere scatterers, and 
the perturbation is a small change in the mass of the moving particle, it 
is straightforward to show that $M(t)$ is equal to the wave-packet 
autocorrelation function evaluated at a scaled time, which can be made to 
be much shorter than the physical time $t$, by choosing a suitably small 
mass perturbation. Therefore {\it for this special perturbation} and hard 
sphere Lorentz gas system, the quantum fidelity can be evaluated for very 
long times, if one knows the behavior of the auto-correlation function for 
a much shorter time interval.  

% ------- To be removed for PRE ------- 
An earlier version of this paper \cite{gd} contained the claim that the
Loschmidt echo, defined above, decays over some time interval as $e^{-2\lambda t}$.
This claim is wrong, and our reasons for reaching this conclusion will be
discussed in Section III.
% ------------------------------------- 

This paper is organized as follows: In Section II, we construct the 
semi-classical propagator for the moving particle and show that when it is 
used to determine the time evolution of a small initial Gaussian wave 
packet, the spreading of the wave packet is, for times less than the 
Ehrenfest time, determined by the classical curvature equations 
\cite{sinai}. These equations describe the rate of spreading of a 
classical, infinitesimal trajectory bundle. In Section III, we apply this 
result to calculate the wave-packet auto-correlation function for periodic 
orbits, as an illustration of the behavior predicted and described by 
Heller \cite{heller}. In Section IV we derive an exact expression for the 
Loschmidt echo, $M(t)$, for a quantum particle moving in a  hard sphere 
Lorentz gas, where the perturbed Hamiltonian is obtained from the 
unperturbed one by a small change of the mass of the moving particle. This 
result allows us to describe the behavior of the echo for long physical 
times in terms of the short time behavior of the wave packet 
auto-correlation function, $C(t)$.
% ------- To be removed for PRE -------
We also discuss the error in our previous paper \cite{gd}. 
% -------------------------------------
The calculations in these sections are 
for two-dimensional systems with hard disk scatterers. The three 
dimensional version of this work is presented in Section V. Here we show 
that the role of the positive Lyapunov exponent in our calculations for two 
dimensional systems is taken by KS entropy, {\it i.e.} by the sum of the 
two positive Lyapunov exponents for the three dimensional system. We 
summarize our results in Section VI.

\section{The semi-classical propagator for the Lorentz gas}
We consider the semi-classical motion in two dimensions, $d=2$, of a 
Gaussian wave packet, with average momentum $\vp_0$, whose initial form is 
given by 
\begin{equation}
\begin{split}
\langle \vr | 0 \rangle \equiv \ps_0(\vr)=&\left(2\pi\sigma_{\|0}\sigma_0
\right)^{-1/2} \\
&\times\exp{\left(\frac{i}{\la}\ze - \frac{\ze^2}{4\Omega_{\|0}} 
- \frac{\et^2}{4\Omega_0} \right)}, 
\end{split}
\label{2.1} 
\end{equation}
where $\la=\hbar/|\vp_0|$ is the de Broglie wavelength of the moving 
particle, $\sigma_{\|0}^2 = 1/\real \: (\Omega_{\|0}^{-1})$ and  
$\sigma_0^2 = 1/\real \: (\Omega_0^{-1})$ characterize the size of the wave 
packet in the $\ze$- and $\et$-directions respectively ($\real\:$ denotes 
the real part). 
\begin{figure}[h] 
\centerline{\epsfig{figure=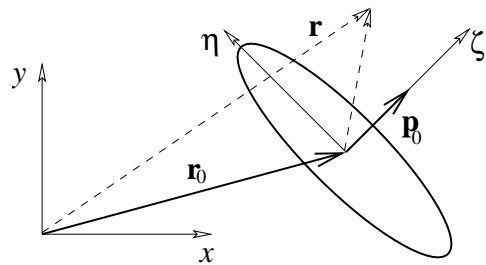,width=2.5in}} 
\caption{Particle-fixed frame of reference at time $t=0$.} 
\label{fig_1} 
\end{figure} 
The $(\ze,\et)$ system of coordinates is chosen with its origin at the 
center of the wave packet, $\vr_0$, and $\ze$-axis pointing in the 
direction of $\vp_0$, with $\et$-axis perpendicular to $\vp_0$:  
\begin{equation}
\vr = \vr_0 + {\bf U}_0 \left( 
\begin{array}{c} 
\ze \\ 
\et 
\end{array} \right), 
\label{2.2} 
\end{equation}
where ${\bf U}_0$ is a $2\times2$ real matrix relating the two coordinate 
systems, see Fig.~\ref{fig_1}. 
 
When the wave packet is far from any scatterers, its time propagation is 
dominated by free streaming, described by the propagator 
\begin{equation}
G_\mathrm{fs}(\vr,\vr',t) = \left( \frac{m}{2 \pi i \hbar t} \right)^{d/2} 
\exp{\frac{i m}{2 \hbar t}(\vr-\vr')^2}, 
\label{2.3} 
\end{equation}
where $m$ is the mass of the moving particle, and $d=2$. Application of 
this propagator to the wave function given by Eq.~(\ref{2.1}) yields, up to 
an irrelevant phase factor, a new Gaussian wave packet of the form of 
Eq.~(\ref{2.1}) with 
\begin{eqnarray}
&\Omega_{\|0} \rightarrow &\Omega_{\|t} = \Omega_{\|0} + i\la vt / 2  
\label{2.4} \\ 
&\Omega_0 \rightarrow &\Omega_t = \Omega_0 + i\la vt / 2, 
\label{2.5} 
\end{eqnarray}
where $v=|\vp_0|/m$ is the average velocity of the particle. The new
particle-fixed frame of reference is related to the stationary one by means 
of Eq.~(\ref{2.2}), with the wave packet center, $\vr_0$, replaced by 
$\vr_t = \vr_0 + (\vp_0/|\vp_0|) vt$ and ${\bf U}_t = {\bf U}_0$. The 
average momentum of the wave packet stays unaffected: $\vp_t = \vp_0$. 
 
To find the semi-classical propagator describing a collision of the 
particle with one scatterer, we start with the general expression for 
the semi-classical propagator as a sum of terms of the form \cite{brbh} 
\begin{equation}
\begin{split}
G_\mathrm{sc}(\vr,\vr',t) =&\left( \frac{1}{2\pi i\hbar} \right)^{d/2} 
|D|^{1/2}\\
&\times\exp{\:i\! \left(\frac{S(\vr,\vr',t)}{\hbar}
+\frac{\pi\mu}{2}\right)},
\end{split}
\label{2.6} 
\end{equation}
where $S(\vr,\vr',t)$ is the classical action along a classical path from 
$\vr'$ to $\vr$ in time $t$, $\mu$ is an index equal to twice the number of 
collisions of the particle with hard disk scatterers over time $t$ 
\cite{gasp}, $D = \det (-\partial^2 S/\partial \vr \partial \vr')$, and 
$d=2$. In general, there are two classical paths connecting points $\vr$ 
and $\vr'$, assuming that $\vr$ is not in the geometric shadow of $\vr'$: a 
reflected path and a direct one. The contribution of the direct path from 
$\vr'$ to $\vr$ to the time evolution of the wave packet is negligible 
after time $t$ if a classical particle with momentum $\vp_0$ would collide 
with the scatterer during the interval $(0,t)$. Thus, we only consider the 
propagator given by the reflected path. 
\begin{figure}[h] 
\centerline{\epsfig{figure=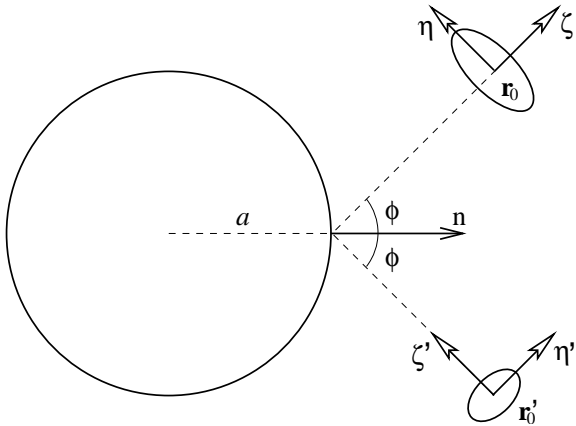,width=3.0in}} 
\caption{Particle-fixed frames of reference: $(\ze',\et')$ at time $0$ and 
$(\ze,\et)$ at $t$.} 
\label{fig_2} 
\end{figure} 
 
Consider a wave packet centered around $\vr'_0$ at time $t=0$ before a 
collision and around $\vr_0$ at $t$ after the collision, Fig.~\ref{fig_2}.
The system of reference originates at the point of classical collision. 
We suppose that the wave-packet to which the propagator will be applied is 
sufficiently small that we only need to find classical trajectories by 
minimizing the action for points starting close to $\vr'_0$ and ending 
close to $\vr_0$ at time $t$. We then write the action, $S_{\bf R}(\vr_0+
\delta\vr,\vr'_0+\delta\vr',t)$, for a trajectory from $\vr'_0+\delta\vr'$, 
$|\delta\vr'| \ll |\vr'_0|$, colliding with a scatterer at point ${\bf R}$ 
and arriving at $\vr_0+\delta\vr$, $|\delta\vr| \ll |\vr_0|$, at time, $t$, 
as 
\begin{equation}
\begin{split}
S_{\bf R}&(\vr_0+\delta\vr,\vr'_0+\delta\vr',t)\\
&=\frac{m}{2t}\left(\left| {\bf R}-\vr'_0-\delta\vr' \right| + \left| \vr_0
+\delta\vr-{\bf R} \right|\right)^2.
\end{split} 
\label{2.7} 
\end{equation}
The variation of this action with respect to the point of collision, 
${\bf R}$, leads to an extremum equation that is used to determine the 
collision point, ${\bf R}$. The algebra simplifies a bit if we take the 
case where $|\vr'_0| =|\vr_0| = r$. Using the variational 
procedure, we find 
\begin{equation}
G_\mathrm{sc}(\ze,\et,\ze',\et',t) = G_\mathrm{sc}^{(1)}(\ze,\ze',t)
G_\mathrm{sc}^{(2)}(\et,\et',t), 
\label{2.8} 
\end{equation}
with 
\begin{equation}
G_{sc}^{(1)}(\ze,\ze',t) \approx \left( \frac{1}{4\pi i \la r} \right)^{1/2}
\exp{\frac{i (\ze + 2r - \ze')^2}{4\la r}} 
\label{2.9} 
\end{equation}
and 
\begin{equation}
\begin{split}
G_{sc}^{(2)}&(\et,\et',t) \approx \left( \frac{a\cos\phi}{4\pi i \la r 
(r + a\cos\phi)} \right)^{1/2}\\
&\times\exp{\frac{i a (\et -\et')^2 \cos\phi + 2i r (\et^2 + (\et')^2)}
{4\la r (r + a\cos\phi)}}.
\end{split}
\label{2.10} 
\end{equation}
Here $\et,\ze$ and $\et',\ze'$ are coordinates with origins at $\vr_0$ and 
$\vr'_0$ respectively, such that $\ze,\ze'$ are along the direction of the 
probability current, and $\et,\et'$ are in directions perpendicular to 
$\ze,\ze'$, respectively, as illustrated in Fig.~\ref{fig_2}. Further, 
$\phi$ is the angle of incidence in the collision. The time dependence in 
the propagator appears in $r$, through the relation $2r=vt$. 
 
There are limits to the range of applicability of the semi-classical 
propagator given by Eq.~(\ref{2.8}). First, the particle's wave function is
supposed to be confined to a small region in space, the linear size of 
which is much smaller that the radius of the scatterer, throughout the time 
interval $(0,t)$. It is this limitation that allows one to consider only 
the reflected path while deriving the propagator. Second, the wave packet 
size is assumed to be much smaller than the distance $r$ between the center 
of the wave packet and the point where the particle would, classically, 
collide with the scatterer. This assumption makes it possible to expand the 
coordinates of points connected by the propagator about corresponding wave 
packet centers.

The propagator in the direction of motion given by Eq.~(\ref{2.9}) is 
simply the free streaming expressed in particle-fixed coordinate frames, 
showing that time evolution of $\ze$-component of the time dependent wave 
packet, $\psi_t$, is unaffected by scattering events. The $\et$-component 
of the propagator, Eq.~(\ref{2.10}), can be easily shown to satisfy the 
identity 
\begin{equation}
\begin{split}
G_\mathrm{sc}^{(2)}(\et,\et',t) = \int d\et_1 &\int d\et_2 \: G_\mathrm{fs}
(\et,\et_2,t/2)\\
&\times \hat{C}(\et_2,\et_1) G_\mathrm{fs}(\et_1,\et',t/2),
\end{split}
\label{2.11} 
\end{equation}
where we introduced an instantaneous collision propagator, 
$\hat{C}(\et,\et')$, according to 
\begin{equation}
\hat{C}(\et,\et') = \delta(\et-\et') \exp \frac{i \et^2}{\la a \cos\phi}. 
\label{2.12} 
\end{equation}
Eq.~(\ref{2.11}) allows to represent the propagator for a  single 
scattering event, $G_\mathrm{sc}(\ze,\et,\ze',\et',t)$, as a product of 
three successive propagators: (i) a free streaming propagator, 
$G_\mathrm{fs}(\ze_1,\et_1,\ze',\et',t/2)$, (ii) an instantaneous collision 
propagator, $\delta(\ze_2-\ze_1)\hat{C}(\et_2,\et_1)$ affecting the 
$\et$-component of $\psi_t$, and (iii) another free streaming propagator, 
$G_\mathrm{fs}(\ze,\et,\ze_2,\et_2,t/2)$. 

Assuming that the wave packet size $\sigma_t$ remains smaller than radius 
$a$ of a scatterer, over the time $t$, we now construct the propagator for 
a trajectory with several collisions of the moving particle with scatterers 
as a combination of free particle and single collision propagators. This is 
appropriate in the semi-classical approximation when the size of the wave 
packet is small compared to the size of a scatterer, and to the average 
separation of the scatterers. Both free flight and instantaneous collision 
propagators leave the Gaussian form of a wave packet invariant. While the 
effect of the free streaming is described by Eqs.~(\ref{2.4}, \ref{2.5}), 
the instantaneous collision propagator, Eq.~(\ref{2.12}), when applied to a 
Gaussian wave packet leads to an instantaneous change in $\Omega$ given by 
\begin{equation}
\frac{1}{\Omega^+} = \frac{1}{\Omega^-} - \frac{4i}{\la a \cos\phi}, 
\label{2.13} 
\end{equation}
where superscripts $\pm$ are used to distinguish variables immediately 
before and immediately after a collision. As mentioned above, $\Omega_\|$ 
is unaffected by instantaneous collisions: $\Omega_\|^+ = \Omega_\|^-$. 
 
The free streaming transformation of $\Omega_t$, coupled with the 
collisional transformation of $\Omega^-$ to $\Omega^+$ given above provides 
a direct connection between this semi-classical analysis of wave packet 
motion and the method of Sinai {\it et al.} for analyzing the ergodic 
properties of the classical Lorentz gas in terms of the curvature of a 
classical wave front \cite{sinai,vbd}. In fact a simple transformation 
allows us to recover the classical equations, and to identify the 
appearance of the positive Lyapunov exponent in the semi-classical 
formulae. To see this let us define complex radii of curvature, $\trho_\|$ 
and $\trho$, according to  
\begin{equation}
\Omega_\| = \frac{i}{2} \la \trho_\| \;\;\;\;\; \mathrm{and} \;\;\;\;\; 
\Omega = \frac{i}{2} \la \trho . 
\label{2.14} 
\end{equation}
In terms of $\trho$, Eqs.~(\ref{2.5} and \ref{2.13}) read 
\begin{eqnarray}
\trho_t = \trho_0 + vt& \;\;\;\;\; &\mathrm{free \:\: streaming,} 
\label{2.15}\\ 
\displaystyle \frac{1}{\trho^+} = \frac{1}{\trho^-} + \frac{2}{a \cos\phi}& 
\;\;\;\;\; &\mathrm{collision,} 
\label{2.16} 
\end{eqnarray}
while $\trho_{\|t} = \trho_{\|0} + vt$ regardless of whether any scattering 
events have taken place over time $t$. These equations for $\trho$ are 
identical with the curvature equations \cite{sinai} for the classical 
Lorentz gas. In an unpublished manuscript describing the diffractive 
scattering of a wave packet by a circular scatterer, Wirzba \cite{wirz} 
noted that the curvature equations can also be extracted from his formalism.

To describe the spreading of a Gaussian wave packet in the Lorentz gas, we 
consider a sequence of collisions parameterized by a set of times 
$\{ t_j \}$ together with a set of collision angles $\{ \phi_j \}$. Direct 
substitution of the free streaming transformation for $\trho_t$, 
Eq.~(\ref{2.15}), into the expression for the size of the wave packet along 
the $\et$-coordinate, {\it i.e.} along the direction perpendicular to the 
average momentum $\vp$ of the particle, $\sigma_t^2 = 1/\real \: 
(\Omega_t^{-1}) = \la/(2 \imag \: (\trho_t^{-1}))$, yields 
\begin{equation}
\sigma_t =  \sigma_{t_j} \left| \frac{\trho_{t_j} + v(t - t_j)}
{\trho_{t_j}} \right| 
= \sigma_{t_j} \exp{\left( v \real \int_{t_j}^t \frac{d\tau}{\trho_\tau} 
\right)}, 
\label{2.17} 
\end{equation}
for $t_j < t < t_{j+1}$. It follows from the relation between $\sigma$ and 
$\trho$, and the change in $\trho$ on collision, that the instantaneous 
scattering transformation does not change the size of the wave packet 
$(\sigma_{t_j}^+ = \sigma_{t_j}^-)$. Thus, we can propagate $\sigma_t$ 
backward in time to get 
\begin{equation}
\sigma_t =  \sigma_0 \exp{\left( v \real \int_0^t \frac{dt'}{\trho_{t'}} 
\right)} = \sigma_0 \: e^{t \lambda_t},  
\label{2.18} 
\end{equation}
where $\sigma_0$ is the initial size of the wave packet at $t = 0$, and  
\begin{equation} 
\lambda_t = \frac{v}{t} \real \int_0^t \frac{d\tau}{\trho_\tau} = 
\frac{v}{t}\int_0^t \frac{d\tau}{\rho_\tau}, 
\label{2.19} 
\end{equation}
where we introduce a {\it real radius of curvature}, $\rho$, as 
\begin{equation}
1/\rho \equiv \real \; (1/\trho). 
\label{2.20} 
\end{equation}
The quantity $\lambda_t$ can be thought of as a wave packet stretching 
exponent over a time $t$. It differs from the classical Lyapunov exponent 
$\lambda$ because it contains quantum effects and the limit of infinite 
time is not taken. The stretching exponent, $\lambda_t$, converges to the 
Lyapunov exponent, $\lambda$, in the long time classical limit: 
\begin{equation}
\lim_{t \rightarrow \infty} \lim_{\la \rightarrow 0} \lambda_t = \lambda. 
\label{2.21} 
\end{equation}  
 
In order to prove Eq.~(\ref{2.21}), one needs to show that $\rho$ becomes 
the classical radius of curvature for the classical Lorentz gas as $\la 
\rightarrow 0$. Substituting Eq.~(\ref{2.20}) along with $\imag \: (1/\trho)
= \la /(2\sigma^2)$ into the transformations for $\trho_t$, 
Eqs.~(\ref{2.15}, \ref{2.16}), one gets 
\begin{equation}
\left. 
\begin{array}{ccl} 
\rho_t &=& \displaystyle \frac{(\rho_0+vt)^2+\varepsilon_0(vt)^2}{\rho_0+vt+
\varepsilon_0 vt} \\ 
&&\\ 
\sigma_t &=& \displaystyle \frac{\sigma_0}{\rho_0} \sqrt{(\rho_0+vt)^2+
\varepsilon_0(vt)^2} 
\end{array} \right\} \;\;\; \mathrm{free \:\: streaming,} 
\label{2.22} 
\end{equation}
and $1/\rho^+ = 1/\rho^- + 2/(a \cos\phi)$ together with $\sigma^+ = 
\sigma^-$ at a collision. Here 
\begin{equation}
\varepsilon_t = \left( \frac{\la\rho_t}{2\sigma_t^2} \right)^2 
\label{2.23} 
\end{equation}
contains all the quantum effects; it vanishes as $\la \rightarrow 0$, which 
makes Eq.~(\ref{2.22}) converge to its classical counterpart 
\cite{sinai,vbd}. Another way to visualize the semi-classical corrections 
is to rewrite Eq.~(\ref{2.22}) in differential form: 
\begin{equation}
\dot{\rho}_t = v(1-\varepsilon_t) \;\;\;\;\; \mathrm{and} \;\;\;\;\; 
\dot{\sigma}_t = v\sigma_t/\rho_t. 
\label{2.24} 
\end{equation}
Here the second equation has its classical form, and the quantum correction 
is apparent in the first equation: it shows that the free flight spreading 
of the wave packet results from a combination of a classical linear 
separation of trajectories and the quantum spreading due to the Uncertainty 
Principle. 
 
The role of the Uncertainty Principle becomes apparent from the following 
simple consideration. Suppose one prepares a tiny minimal wave packet with 
spatial uncertainty $\sigma_0$. The corresponding uncertainty in momentum, 
$\Delta p$, is then given by $\sigma_0 \Delta p = \hbar/2$. After some time 
$t$ the wave packet size evolves to $\sigma_\mathrm{UP} \approx 
(\Delta p/m)t = \la v t/(2\sigma_0)$ merely due to the Uncertainty 
Principle. Writing the geometrical (classical) spreading as 
$\sigma_\mathrm{CL} = \sigma_0 (1 + vt/\rho_0)$, we notice that $\sigma_t$ 
in Eq.~(\ref{2.22}) is essentially a simple combination of 
$\sigma_\mathrm{CL}$ and $\sigma_\mathrm{UP}$, namely $\sigma_t = 
\sqrt{\sigma_\mathrm{CL}^2 + \sigma_\mathrm{UP}^2}$.  
 
\begin{figure}[h] 
\centerline{\epsfig{figure=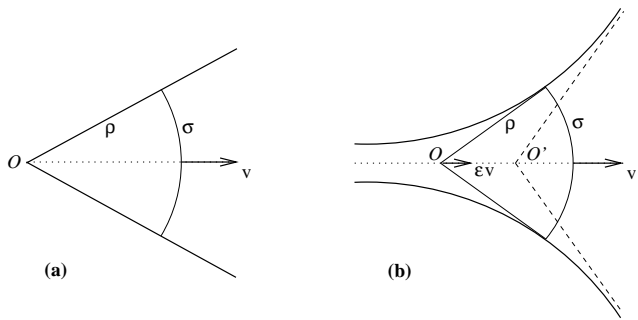,width=3.3in}} 
\caption{Free flight time evolution of $\rho$ and $\sigma$: {\bf (a)} 
classical case, $\varepsilon = 0$, {\bf (b)} quantum case, 
$\varepsilon > 0$.} 
\label{fig_3} 
\end{figure} 
 
Fig.~\ref{fig_3} illustrates the free flight dynamics of $\rho$ and 
$\sigma$ given by Eq.~(\ref{2.24}). Fig.~\ref{fig_3}(a) pictures the 
classical limit, $\la=0$: an arc of instantaneous radius $\rho$ and length 
$\sigma$ moves with constant velocity $v$ along the ``cone'' originating at 
a point $O$. Eq.~(\ref{2.24}) with $\varepsilon_t=0$ describes the time 
evolution of $\rho$ and $\sigma$ in this case. In quantum regime, 
$\varepsilon > 0$, the point $O$ is also moving in the same direction as 
the arc, but with a different, time dependent, velocity equal to 
$\varepsilon v$, see Fig.~\ref{fig_3}(b). It can be shown from 
Eqs.~(\ref{2.23}, \ref{2.24}) that $\varepsilon_t \sim t^{-2}$ as 
$t\rightarrow\infty$, implying the convergence of point $O$ to some point 
$O'$ in the long time limit, see Fig.~\ref{fig_3}(b). The time evolution of 
$\rho$ and $\sigma$ is then dominated by classical equations when $O$ is 
close to $O'$.  
 
\begin{figure}[h] 
\centerline{\epsfig{figure=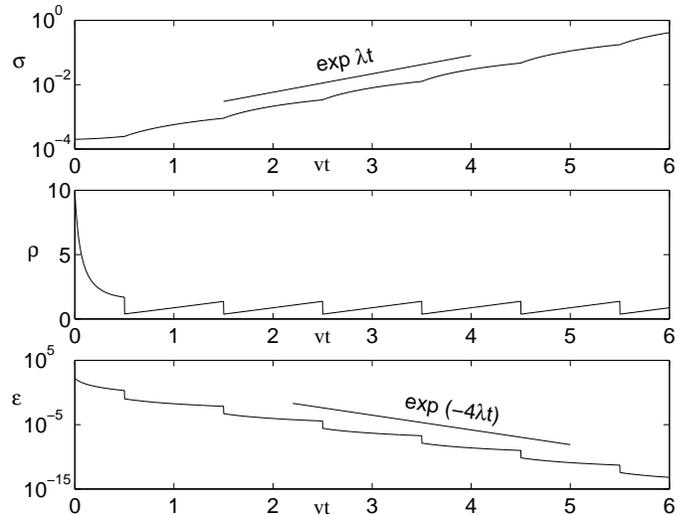,width=3.5in}} 
\caption{Wave packet size, $\sigma$, real radius of curvature, $\rho$, and 
$\varepsilon$, are shown as functions of time, $t$, for a two-disk periodic 
orbit. Disk radii $a=1$, center-to-center separation $R=3$, de Broglie wave 
length $\la=10^{-7}$. The corresponding Lyapunov exponent $\lambda/v 
\approx 1.32$. Initial wave packet size $\sigma_0=2\cdot10^{-4}$ and 
$\rho_0=10$. The particle is located in the middle between the two disks at 
$t=0$. Exponential trends are shown for plots of $\sigma$ and 
$\varepsilon$. All distances are measured in units of $a$.}
\label{fig_4} 
\end{figure}

We now define an interval of time, called the {\it Lyapunov regime} for 
which the values of $\rho$ and $\sigma$ satisfy the inequality $\varepsilon 
\ll 1$. It follows from the free flight and collision transformations for 
$\rho_t$ and $\sigma_t$ that $\varepsilon_t$ is a rapidly decreasing 
function of time, see Fig.~\ref{fig_4}. Therefore, once in the Lyapunov 
regime the system stays in it for some time $t_L$, at which $\sigma$ 
becomes comparable with the size of scatterers, and our collision analysis 
breaks down. It can also be shown that if the Lyapunov regime inequality is 
not satisfied at $t=0$, and the wave packet is small, the system rapidly 
evolves to a state for which this inequality is satisfied. During this 
transient regime $\rho$ rapidly decreases whereas $\sigma$ does not change 
significantly, see Fig.~\ref{fig_4}. 
  
In the Lyapunov regime, the first equation in Eqs.~(\ref{2.24}) reduces to 
its classical counterpart, $\dot{\rho} \approx v$, so that $\sigma_t$ grows 
exponentially in the same way as a small pencil of trajectories separates 
exponentially in the classical system. That is $\sigma_t =\sigma_0 
\exp{(t\lambda_t)}$, where $\lambda_t$ is given by Eq.~(\ref{2.19}) and 
calculated using only classical mechanics. It is useful to remark that 
$\lambda_t$ typically reaches a value close to the classical Lyapunov 
exponent $\lambda$ after only a few collisions, see Fig.~\ref{fig_4}. On 
the other hand, one can estimate that the upper limit of $t_L$, the maximum 
duration of the Lyapunov regime, is $t_L^{\mathrm max} \sim (1/(2\lambda)) 
\ln (a/\la)$, that is, about half of the Ehrenfest time \cite{haake}, which 
for sufficiently small $\la$ can be long enough for the wave packet to 
exhibit exponential spreading. 
 
Finally, we illustrate the exponential spreading of a Gaussian wave packet 
for the case of particles moving in short, periodic orbits. We numerically 
evaluate $\sigma_t$ and $\rho_t$ in Eq.~(\ref{2.24}) (and in the $\rho^-$ 
to $\rho^+$ collision transformation) for the simplest periodic orbit: a 
particle moving back and forth along the line connecting the centers of two 
disks. Fig.~\ref{fig_4} shows $\sigma_t$, $\rho_t$, and quantity 
$\varepsilon_t$, given by Eq.~(\ref{2.23}), for the two disks of radius 
$a=1$, and the center-to-center separation $R=3$. The particle is placed 
in the middle between the two disks at $t=0$, and has the de Broglie 
wavelength $\la = 10^{-7}$. The initial wave packet is characterized by 
$\sigma_0 = 2\cdot10^{-4}$ and $\rho_0 = 10$, so that $\varepsilon_0 
\approx 156$ and the system is far from the Lyapunov regime at $t=0$. 
Fig.~\ref{fig_4} shows that it only takes a single collision for the system 
to reach the Lyapunov regime, $\varepsilon \ll 1$.
 
The parameters in Fig.~\ref{fig_4} are chosen so as to illustrate the 
essential regimes: a short decay of quantum effects ($\varepsilon$ becomes 
less than unity), followed by the Lyapunov spreading of the wave packet, 
$\sigma \sim \exp (\lambda t)$. The relatively small value of the de 
Broglie wavelength used in this example can be indeed achieved 
experimentally \cite{exper}.  
 
The classical Lyapunov exponent of a two-disk periodic orbit is known 
exactly \cite{jose,gaspbk}, 
\begin{equation}
\lambda = \frac{v}{R-2a}\ln\frac{R-a+\sqrt{R(R-2a)}}{a}. 
\label{2.25} 
\end{equation}
In our case Eq.~(\ref{2.25}) gives $\lambda/v \approx 1.32$. The numerical 
evaluation presented in Fig.~\ref{fig_4} shows that a single collision is 
enough to initiate the exponential growth of the wave packet (with the rate 
given by the classical Lyapunov exponent), which persists for about 5-6 
collisions. Our results do not apply for times longer than the duration of 
the Lyapunov regime, $vt_L/a \approx 6$.

\section{The wave-packet auto-correlation function} 
As an application of the analysis of wave packet dynamics developed above, 
we calculate the wave packet auto-correlation function, $C(t)$, defined in 
Eq.~(\ref{1.1}), for particles moving in periodic orbits.  Here the initial 
state, $|0\rangle$, describes a Gaussian wave packet centered about $\vr_0$ 
with its average momentum $\vp_0$, such that the phase point $(\vr_0, 
\vp_0)$ lies on a periodic orbit of the corresponding classical system. The 
auto-correlation function for periodic orbits in billiard systems was 
studied by Heller \cite{heller} some time ago  using different techniques. 
The calculations presented here agree with Heller's results and provide 
some additional information about this correlation function.
 
The reasons for restricting our calculations to periodic orbits are as 
follows. The expansion used above to obtain the semi-classical single 
collision propagator in the previous section, Eq.~(\ref{2.8}), is correct 
for wave packets which are small compared to disk radii and average 
separation among scatterers. Mathematically, this limitation is a 
consequence of the truncation of the expansion of the coordinates of 
starting and final points connected by the propagator, $\vr'$ and $\vr$ 
respectively, about the centers of initial and final wave packets 
respectively, {\it i.e.} $\vr' = \vr'_0 + \delta\vr'$ and $\vr = \vr_0 
+ \delta\vr$. Therefore, one gets a close approximation to the particle's 
wave function at positions close to the wave packet center, $\vr_0$, but 
the approximation may fail on the periphery of the wave packet. Our 
calculations of the auto-correlation function, $C(t)$, or Loschmidt echo, 
$M(t)$, are only reliable when the relevant overlap integrals are dominated 
by central region of the wave function, and contributions coming from wave 
packet wings can be neglected. This condition is most easily satisfied when 
the classical motion is along a periodic orbit.
% ------- To be removed for PRE -------
The result for the decay of the Loschmidt echo reported earlier \cite{gd},
as $\exp(-2\lambda t)$, is incorrect because this restriction on the validity
of the semi-classical approximation was not properly taken into account. This
error first became apparent when we obtained an exact result that is inconsistent
with this decay. The exact result is described in the next section.
% ------------------------------------- 
  
\begin{figure}[h] 
\centerline{\epsfig{figure=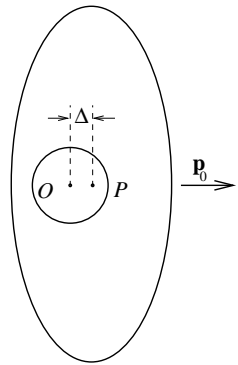,width=1.2in}} 
\caption{A Gaussian wave packet is shown at time $t=0$ (centered about 
point $O$), and at a later time $t=nT + \Delta/v$ (centered about point 
$P$). Points $O$ and $P$ lie on the same periodic orbit, and are separated 
in time by $n$ ($=0,1,2,\ldots, N$) periods, $T$, of the periodic orbit, 
plus a short time interval $\Delta/v$. The separation distance $\Delta$ is 
assumed to be sufficiently small in order for the initial and final wave 
packets to overlap significantly.} 
\label{fig_5} 
\end{figure} 
 
Consider a wave packet whose initial average coordinate, $\vr_0$, and 
momentum, $\vp_0$, correspond to a phase space point on a periodic orbit, 
with period $T$, of the classical Lorentz gas. Suppose that $nT$ is smaller 
than the Ehrenfest time, for $n = 0, 1, 2, \ldots, N$ so that we can apply 
the analysis developed in the previous section in order to propagate the 
wave packet over times 
\begin{equation}
t = nT + \Delta / v, 
\label{3.1} 
\end{equation}
where the displacement $\Delta$ is sufficiently small in order for the 
initial and final wave packets to overlap significantly, as illustrated in 
Fig.~\ref{fig_5}. For simplicity we take the initial wave function, 
$\psi_0$, to be a ``circular'' wave packet, {\it i.e.} $\sigma_{\|0} = 
\sigma_0$ and $\rho_{\|0} = \rho_0$ with the explicit form, 
\begin{equation}
\begin{split}
\psi_0&(x, y) = \left( \frac{1}{2\pi\sigma_0^2} \right)^{1/2}\\
&\times\exp\left[ -\frac{1}{4}\left( \frac{1}{\sigma_0^2} - 
\frac{2i}{\la\rho_0} \right) (x^2 + y^2) + \frac{i}{\la} x \right], 
\end{split}
\label{3.2} 
\end{equation}
where $x$-axis is directed along $\vp_0$. In the same coordinate system the 
wave packet $\psi_t$ propagated from $\psi_0$ over time $t$, given by 
Eq.~(\ref{3.1}), reads, up to an irrelevant phase factor, 
\begin{equation}
\begin{split}
\psi_t(x, y) = &\left( \frac{1}{2\pi\sigma_{\|}\sigma} \right)^{1/2} 
\exp\left[ -\frac{1}{4}\left( \frac{1}{\sigma_\|^2} - \frac{2i}{\la\rho_\|} 
\right) (x-\Delta)^2 \right.\\
&\left. - \frac{1}{4}\left( \frac{1}{\sigma^2} - \frac{2i}{\la\rho} \right) 
y^2 + \frac{i}{\la} (x-\Delta) \right]. 
\end{split}
\label{3.3} 
\end{equation}
Here, $\sigma_\|$, $\sigma$, $\rho_\|$ and $\rho$ depend on time $t$ 
through a sequence of free flight and collision transformations developed 
in the previous section. The probability distribution $|\psi_0(\vr)|^2$ is 
negligible outside a small circle of radius $r \sim \sigma_0$. Therefore, 
the main contribution to the overlap $\langle \psi_0 | \psi_t \rangle$ 
comes from the points inside this circle, and the central regions of the 
wave packets will dominate the integrals for small center-to-center 
separations $\Delta$, as illustrated in Fig.~\ref{fig_5}. 
 
\begin{figure}[h] 
\centerline{\epsfig{figure=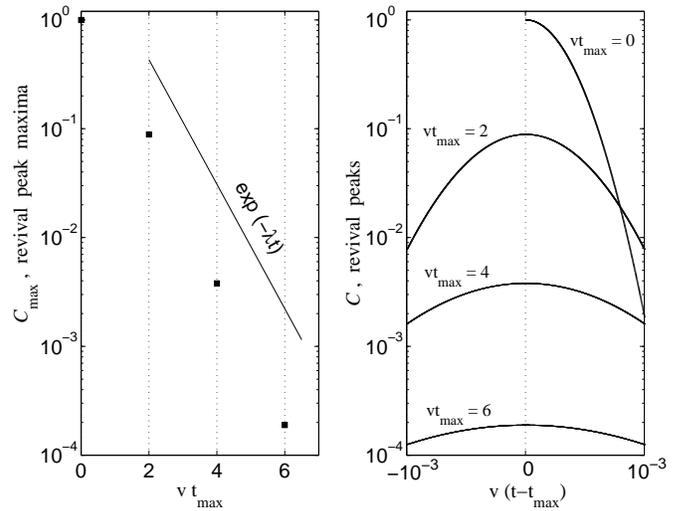,width=3.5in}} 
\caption{Revival peaks of the wave packet auto-correlation function, 
$C(t)$, for the same two-disk periodic orbit as in Fig.~\ref{fig_4}: $a=1$ 
and $R=3$. Particles de Broglie wave length $\la=10^{-7}$. The initial wave 
packet is characterized by $\sigma_0=\sigma_{\|0}=2\cdot10^{-4}$ and 
$\rho_0=\rho_{\|0}=10$. The exponential trend is indicated by a straight 
line.} 
\label{fig_6} 
\end{figure}

A straightforward calculation shows that for $t$ given by Eq.~(\ref{3.1})
\begin{equation}
C(t) = \left| \int d\vr \: \psi_0^*(\vr) \psi_t(\vr) \right|^2 = 
\frac{A}{\sigma} \exp(-\alpha\Delta^2), 
\label{3.4} 
\end{equation}
where 
\begin{eqnarray}
A &=& \frac{4}{\sigma_0^2} \left| \frac{g_\| g}{\sigma_\|} \right|, 
\label{3.5} \\ 
\alpha &=& \frac{1}{2} \real \left[ g_\| \left( \frac{1}{\sigma_\|^2} - 
\frac{2i}{\la\rho_\|} \right) \left( \frac{1}{\sigma_0^2} + 
\frac{2i}{\la\rho_0} \right) \right], 
\label{3.6} 
\end{eqnarray}
with 
\begin{equation}
\begin{split}
g &\equiv \left[ \frac{1}{\sigma^2} + \frac{1}{\sigma_0^2} - \frac{2i}{\la} 
\left( \frac{1}{\rho} - \frac{1}{\rho_0} \right) \right]^{-1},\\
g_\| &\equiv \left[ \frac{1}{\sigma_\|^2} + \frac{1}{\sigma_0^2} - 
\frac{2i}{\la} \left( \frac{1}{\rho_\|} - \frac{1}{\rho_0} \right) 
\right]^{-1}. 
\end{split}
\label{3.7} 
\end{equation}
As seen from Eq.~(\ref{3.4}) the auto-correlation function exhibits a 
sequence of peaks corresponding to partial reconstruction of the wave 
packet at times $t = nT$. These peaks, first studied by Heller 
\cite{heller}, have a simple physical origin: the wave packet repeatedly 
passes through the starting point giving rise to strong maxima of the 
return probability $C(t)$. These maxima are periodic orbit revivals and 
should be distinguished from more general classes of quantum revivals that 
do not require a particular periodic orbit for their appearance 
\cite{toms}. It can be shown that time dependence of $A$ and $\alpha$ is
sub-exponential compared to to the exponential growth of $\sigma$ with 
time, so that the periodic orbit revival peaks have predominantly Gaussian 
form. The strength of the peaks decreases exponentially with time with a
rate given by the Lyapunov exponent of the periodic orbit. This follows 
from the fact that height of the peaks is mainly determined by the 
exponential growth of the size of a wave packet, $\sigma \sim 
\exp(\lambda t)$. It is worth noting that $A$ decays with time in a power 
law manner making the auto-correlation function to decay slightly faster 
than $\exp(-\lambda t)$, see Fig.~\ref{fig_6}.
 
Fig.~\ref{fig_6} shows the numerical evaluation of the revivals in 
Eq.~(\ref{3.4}) for the two-disk periodic orbit described in previous 
section, see Fig.~\ref{fig_4}. A particle of the de Broglie wavelength 
$\la=10^{-7}$ moves back and forth between two disks of radii $a=1$, with 
the center-to-center separation $R=3$, along the line connecting the 
centers. The initial wave packet is located in the middle between the two 
disks, and is characterized by $\sigma_{\|0}=\sigma_0=2\cdot10^{-4}$ and 
$\rho_{\|0}=\rho_0=10$. The left part of Fig.~\ref{fig_6} shows the revival 
maxima $C_\mathrm{max}$, which occur at $t_\mathrm{max}=nT$. The right part 
shows the auto-correlation function in small neighborhoods of the 
corresponding maxima. 
 
It easy to show that in case of the two-disk periodic orbit, the revival 
peaks are separated by deep minima of the auto-correlation function. The 
minima occur when the average momenta of the original wave packet and the 
one propagated in time $t$ are pointing in opposite directions. When this 
happens, $\psi_0$ and $\psi_t$ interfere destructively, and the 
auto-correlation integral, $C(t)$, is very small. We calculate the overlap 
of $\psi_0$ and $\psi_t$ at $t=(n+1/2)T$, when the two wave packets are 
centered about the same point, but move in opposite directions. The initial 
wave function is given by Eq.~(\ref{3.2}), while 
\begin{equation}
\begin{split}
\psi_t(x,y) = &\left( \frac{1}{2\pi\sigma_{\|}\sigma} \right)^{1/2} 
\exp\left[ -\frac{1}{4}\left( \frac{1}{\sigma_\|^2} - \frac{2i}{\la\rho_\|} 
\right) x^2 \right.\\
&\phantom{aaaaaaaaa}\left. -\frac{1}{4}\left( \frac{1}{\sigma^2} - 
\frac{2i}{\la\rho} \right) y^2 - \frac{i}{\la} x \right]. 
\label{3.8} 
\end{split}
\end{equation}
Then, 
\begin{equation}
C(t) = \left| \int d\vr \: \psi_0^*(\vr) \psi_t(\vr) \right|^2 = 
\frac{A}{\sigma} \exp \left( -\frac{2}{\la^2} \real \; g_\| \right), 
\label{3.9} 
\end{equation}
where $A$ and $g_\|$ are defined in Eqs.~(\ref{3.5}, \ref{3.7}). It can be 
shown that the exponential in Eq.~(\ref{3.9}) is a very small number if the 
condition $\la \ll \sigma_0, \rho_0$ is satisfied, {\it e.g.} in case of 
the periodic orbit considered above (Fig.~\ref{fig_4} and Fig.~\ref{fig_6}) 
$\real \; g_\|$ ranges from $2\cdot10^{-8}$ to $4\cdot10^{-8}$ making 
$C(t)$ smaller than $(A/\sigma) \exp(-4\cdot10^6)$, which is practically 
zero when compared with values of $C(t)$ at the PO revival maxima.  
 
Similar deep minima of the auto-correlation function will occur for 
particles in more complicated periodic orbits, making the revivals very 
pronounced.

\section{A simple hard sphere Loschmidt echo} 
Having described the wave-packet auto-correlation function for the Lorentz 
gas, we now present a simple identity that allows us to use this 
correlation function to calculate, analytically, the Loschmidt echo for a 
{\it very particular perturbation}. The Loschmidt echo was defined by 
Eq.~(\ref{1.2}). We suppose that the perturbed Hamiltonian is obtained from 
the unperturbed one by changing the mass of the moving particle, from $m$ 
to $m+\delta m$. The identity depends on the fact that for hard scatterers, 
no matter what their shape, the eigenfunctions depend on wave numbers 
rather than on the mass of the moving particle. That is, the sets of 
eigenfunctions for particles of different masses are the same, only the 
values of the energy corresponding to the same wave numbers differ. The 
wave functions are the solutions of the scalar Helmholtz equation 
\begin{equation}
\left(\nabla^2 + k^2 \right)\phi_\vk(\vr)=0 
\label{she} 
\end{equation}
which satisfy the Dirichlet boundary condition that $\phi_\vk(\vr)$ 
vanishes on the surface of each scatterer, and on the boundaries of the 
system. 
 
We can express the time propagator for a moving particle of mass $m$ under 
a Hamiltonian $H_m$ as 
\begin{equation} 
\exp \left( -\frac{it}{\hbar}H_m \right) = \sum_\vk | \phi_\vk \rangle 
\exp\left( -\frac{i\hbar t}{2m} k^2 \right) \langle \phi_\vk |, 
\label{4.1} 
\end{equation}
where the summation is over all possible eigenstates of the system. These 
eigenstates, $| \phi_\vk \rangle$, satisfy an orthonormality relation 
\begin{equation}
\langle \phi_{\vk'} | \phi_\vk \rangle = \delta_{\vk',\vk}, 
\label{4.2} 
\end{equation}
where the choice between Kronecker and Dirac delta functions is dictated by 
the nature of the eigenstates. Eqs.~(\ref{4.1}, \ref{4.2}) hold for systems 
with hard wall potentials in any number of spatial dimensions. 
 
This representation of the time displacement operator, Eq.~(\ref{4.1}), 
together with the orthonormality condition, Eq.~(\ref{4.2}), leads to the 
following identity 
\begin{equation}
\exp \left( \frac{it}{\hbar}H_{m + \delta m} \right) \exp \left( -
\frac{it}{\hbar}H_m \right) = \exp \left( -\frac{it_s}{\hbar}H_m \right), 
\label{4.3} 
\end{equation}
where $t_s$ is a scaled time, related to the physical time $t$ by 
\begin{equation}
t_s = \frac{\delta m}{m + \delta m}t. 
\label{4.4} 
\end{equation}
  
This identity permits us to express the Loschmidt echo, for this special 
perturbation, in terms of the wave-packet auto-correlation function as 
\begin{equation}
\begin{split} 
M(t)= \left| \langle 0 | \exp \left( \frac{it}{\hbar}H_{m + \delta m} 
\right) \exp \left( -\frac{it}{\hbar}H_m \right) | 0 \rangle \right|^2& \\ 
= \left| \langle 0 | \exp \left( -\frac{it_s}{\hbar}H_m \right) | 0 \rangle 
\right|^2 = C(t_s)&. 
\end{split} 
\label{4.5} 
\end{equation}
This is the main result of this section. For small perturbations, 
$\delta m/m \ll 1$, the Loschmidt echo for long times can be expressed in 
terms of the short time auto-correlation function. 

The physical origin of this result is straightforward. Classically, the 
perturbed and unperturbed masses follow exactly the same trajectory, but 
with different velocities. Hence the forward motion with mass $m$ followed 
by the reversed motion with mass $m+\delta m$ has a final position which is 
different from the initial position, and corresponds to motion over the 
part of the path that is not reached by the time reversed trajectory. This 
is exactly reflected by the operator identity, Eq.~(\ref{4.3}). Perhaps the 
most remarkable thing about this result is the fact that although for long 
times a small wave packet will spread over large distances, the Loschmidt 
echo in this case is determined by the short time spreading of the wave 
packet, even if the physical time is quite large. 
% ------- To be removed for PRE -------
As mentioned in Section III, the exact result, Eq.~(\ref{4.5}), is
inconsistent with the Lyapunov decay reported earlier \cite{gd}. 
% ------------------------------------- 

As discussed in Section III, the auto-correlation function $C(t)$ exhibits 
a sequence of sharp revival maxima when the particle moves on a classically 
periodic orbit. The maxima occur at times $t_\mathrm{max}$ multiple of the 
period $T$ of the periodic orbit, and $C(t_\mathrm{max}) \sim \exp(-\lambda 
t_\mathrm{max})$, where $\lambda$ is the corresponding classical Lyapunov 
exponent. According to Eq.~(\ref{4.5}) the mass perturbation Loschmidt echo 
$M$ at time $t$ is simply the auto-correlation function $C$ at the scaled 
time $t_s$ given by Eq.~(\ref{4.4}). Thus like the auto-correlation 
function, the Loschmidt echo $M(t)$ exhibits a periodic sequence of maxima 
at times 
\begin{equation} 
t'_\mathrm{max} = \frac{m + \delta m}{\delta m} t_\mathrm{max}  = n 
\frac{m + \delta m}{\delta m} T, 
\label{4.6} 
\end{equation} 
where $n=1,2,\ldots,N$, such that $NT$ is smaller than the duration of the 
Lyapunov regime, $t_L$. The envelope of the maxima exhibits a mainly 
exponential decay: 
\begin{equation} 
M(t'_\mathrm{max}) = C(t_\mathrm{max}) \sim \exp(-\lambda t_\mathrm{max}) = 
\exp(-\lambda_s t'_\mathrm{max}). 
\label{4.7} 
\end{equation} 
Here we introduced a {\it scaled} Lyapunov exponent according to 
\begin{equation} 
\lambda_s = \frac{\delta m}{m + \delta m} \lambda. 
\label{4.8} 
\end{equation} 
It is important to note that the behavior of the Loschmidt echo described 
by Eq.~(\ref{4.7}) can persist for times much longer than $t_L$ (for 
sufficiently small $\delta m$) despite the fact that the analysis of the 
wave packet dynamics presented in Section II is valid only for times 
shorter than $t_L$. 

The Hamiltonian perturbation used in this section is rather trivial since 
the perturbed Hamiltonian commutes with the unperturbed Hamiltonian. 
Therefore the results of this section are not to be compared with those 
obtained for more complicated perturbations such as distortion of the mass 
tensor \cite{past,cucc}.

\section{Generalization to three dimensions} 
The derivation of the wave packet propagator in Section II, and the 
calculation of the periodic orbit revivals, Section III, were carried out 
for hard-disk systems in two dimensions, $d=2$. We now generalize these 
calculations to the three dimensional case, $d=3$, using methods similar to 
those used to describe the classical separation of close trajectories 
\cite{sinai,vbd} 
 
The initial Gaussian wave packet in three dimensions reads 
\begin{equation}
\begin{split}
\langle \vr | 0 \rangle \equiv \ps_0&(\vr) = 
\left( \frac{1}{2\pi}\right)^{3/4} \left( \frac{1}{\sigma_{\|0} \sigma_0^2} 
\right)^{1/2}\\
&\times\exp\left( \frac{i}{\la}\ze - \frac{\ze^2}{4\Omega_{\|0}} - 
\frac{1}{4} \vet^\T \vOmega_0^{-1} \vet \right), 
\end{split}
\label{5.1} 
\end{equation}
where $\ze$-axis is directed along the momentum $\vp_0$, see 
Fig.~\ref{fig_1}, while $\vet\equiv(\et^{(1)}, \et^{(2)})^\T$ lies in the 
plane perpendicular to $\vp_0$; $\vOmega_0$ is a $2\times2$ complex 
symmetric matrix, and $\T$-superscript denotes transposition. As in 
two-dimensional case, the origin of the orthogonal frame $(\ze, \et^{(1)}, 
\et^{(2)})$ travels with the center of the wavepacket with fixed axes, 
except at collisions, when the axes rotate so that the new $\zeta$ axis is 
in the direction of motion of the center of the wave packet, see 
Fig.~\ref{fig_1}. The wave packet size in $\ze$-direction $\sigma_{\|0}^2 = 
1/\real \: (\Omega_{\|0}^{-1})$, while in $\vet$-plane 
\begin{equation}
\sigma_0^2 = \frac{1}{\sqrt{\det\real \:(\vOmega_0^{-1})}}. 
\label{5.2} 
\end{equation}
 
Application of the free streaming propagator $G_\mathrm{fs}(\vr,\vr',t)$, 
given by Eq.~(\ref{2.3}) with $d=3$, to the wave function above changes 
$\vOmega_0$ to 
\begin{equation}
\vOmega_t = \vOmega_0 + i(\la vt / 2)\unit, 
\label{5.3} 
\end{equation}
where $\unit$ is the $2\times2$ unit matrix; the change of the 
$\ze$-directional component of the wave packet is the same as in the 
two-dimensional case, Eq.~(\ref{2.4}). 
 
The single-sphere scattering propagator is given by Eq.~(\ref{2.6}) with 
$d=3$. As in the two-dimensional problem, only the reflected path 
contributes to the propagator for a wave packet small compared to the 
sphere radius, $a$. Closely following the arguments of Section II in three 
dimensions, one can verify that the scattering propagator 
$G_\mathrm{sc}(\vr,\vr',t)$ can be written as 
\begin{equation}
\begin{split}
G_\mathrm{sc}(\vr,\vr',t) = \int d\vr_1 &\int d\vr_2 \: 
G_\mathrm{fs}(\vr,\vr_2,t/2)\\
&\times\hat{C}(\vr_2,\vr_1) G_\mathrm{fs}(\vr_1,\vr',t/2), 
\end{split}
\label{5.4} 
\end{equation}
where, in order to simplify the algebra, we consider the case that the 
corresponding classical collision takes place at time $t/2$. The 
instantaneous collision transformation $\hat{C}$, when expressed in 
particle-fixed coordinate frames $(\ze_1,\et_1^{(1)},\et_1^{(2)})$ and 
$(\ze_2,\et_2^{(1)},\et_2^{(2)})$ just before and after the collision 
respectively, reads 
\begin{equation}
\begin{split}
\hat{C}(\ze_2,\vet_2,\ze_1,\vet_1) = \delta&(\ze_2-\ze_1)\delta(\vet_2-
\vet_1)\\
&\times\exp \frac{i}{\la a} \vet_1^\T \vQ(\phi,\theta) \vet_1, 
\end{split}
\label{5.5} 
\end{equation}
where 
\begin{equation}
\vQ(\phi,\theta) = \vP_\theta \: \mathrm{diag}\left[\frac{1}{\cos\phi}, 
\cos\phi\right] \vP_\theta^\T, 
\label{5.6} 
\end{equation}
and 
\begin{equation}
\vP_\theta = \left(  
\begin{array}{rr} 
\cos\theta & -\sin\theta \\ 
\sin\theta & \cos\theta 
\end{array} \right). 
\label{5.7} 
\end{equation}
Here $\phi$ is the angle of incidence in the collision plane, see 
Fig.~\ref{fig_2}, and $\theta$ is the azimuthal angle that 
$\et_1^{(1)}$-axis makes with the collision plane. Note, that the 
coordinate frames $(\ze_1,\et_1^{(1)},\et_1^{(2)})$ and $(\ze_2,\et_2^{(1)},
\et_2^{(2)})$ are related to each other by the $3\times3$ reflection matrix 
$(\unit_3 - 2{\bf n}{\bf n})$, where $\unit_3$ is the $3\times3$ unit 
matrix, and ${\bf n}$ stands for the three-dimensional collision vector, as 
illustrated in Fig.~\ref{fig_2}. 
 
As seen from Eq.~(\ref{5.5}) the instantaneous collision does not affect 
$\Omega_\|$, but changes the $\vet$-component of the wave packet according 
to 
\begin{equation}
\vOmega^{-1(+)} = \vOmega^{-1(-)} - \frac{4i}{\la a} \vQ(\phi,\theta). 
\label{5.8} 
\end{equation}
Introducing the radius of curvature matrix $\vtrho$ as $\vOmega \equiv 
i\la\vtrho/2$ we obtain the three-dimensional equivalent of 
Eqs.~(\ref{2.15}, \ref{2.16}), 
\begin{eqnarray}
\vtrho_t = \vtrho_0 + vt \unit & &\mathrm{free \:\: streaming,} 
\label{5.9}\\ 
\vtrho^{-1(+)} = \vtrho^{-1(-)} + \frac{2}{a} \: \vQ(\phi,\theta) & 
&\mathrm{collision.} 
\label{5.10} 
\end{eqnarray}
Both transformations preserve the symmetry of the complex matrix $\vtrho$. 
 
As in two-dimensional case, we consider a sequence of collisions 
parameterized by a set of times $\{ t_j \}$ together with a set of 
collision angles $\{ \phi_j, \theta_j \}$. Substitution of the free 
streaming transformation, Eq.~(\ref{5.9}), into the expression for the size 
of the wave packet in the $\vet$-plane, $\sigma_t^2 = 
\la/(2\sqrt{\det\imag\:\vtrho_t^{-1}})$, yields 
\begin{equation}
\begin{split}
\sigma_t^2 = \sigma_{t_j}^2 &\left| \frac{\det\left[\vtrho_{t_j} + 
v(t - t_j)\unit\right]}{\det\vtrho_{t_j}} \right|\\
&= \sigma_{t_j}^2 \exp{\left( v \real \int_{t_j}^t d\tau \: 
\mathrm{tr}\vtrho_\tau^{-1} \right)}, 
\end{split}
\label{5.11} 
\end{equation}
for $t_j < t < t_{j+1}$. Here we used the identity $\det\imag\:
\vtrho^{-1}=|\det\vtrho|^{-2}\det\imag\:\vtrho$. By propagating $\sigma_t$ 
backward in time we find 
\begin{equation}
\sigma_t^2 =  \sigma_0^2 \exp\left( v \real \int_0^t d\tau \: 
\mathrm{tr}\vtrho_\tau^{-1} \right) = \sigma_0^2 \: e^{t h_t},  
\label{5.12} 
\end{equation}
where $\sigma_0$ characterizes the wave packet at $t = 0$, and  
\begin{equation} 
h_t = \frac{v}{t} \real \int_0^t d\tau \: \mathrm{tr}\vtrho_\tau^{-1}. 
\label{5.13} 
\end{equation}
The quantity $h_t$ is a time-dependent stretching exponent, which describes 
growth of the area of wave packet cross section perpendicular to the 
direction of particle's motion. It can be shown to converge in the long 
time classical limit to the classical Kolmogorov-Sinai (KS) entropy 
$h_\mathrm{KS}$, equal to the sum of all positive Lyapunov exponents in the 
system: 
\begin{equation}
\lim_{t \rightarrow \infty} \lim_{\la \rightarrow 0} h_t = h_\mathrm{KS} = 
\sum_{\lambda_j>0} \lambda_j. 
\label{5.14} 
\end{equation}  
 
To complete the analogy with the two-dimensional problem we define a {\it 
real} radius of curvature matrix $\vrho$ and a {\it real} $2\times2$ matrix 
$\vSig$ in accordance with 
\begin{equation}
\vtrho \equiv \left[ \vrho^{-1} + \frac{i\la}{2} (\vSig\vSig^\T)^{-1} 
\right]^{-1}. 
\label{5.15} 
\end{equation}
It is easy to show that $\vSig$ determines the size $\sigma$ of the wave 
packet, $\sigma^2=|\det\vSig|$, and is not affected by the collision 
transformation given by Eq.~(\ref{5.10}), while $\vrho$ satisfies 
\begin{equation}
\vrho^{-1(+)} = \vrho^{-1(-)} + \frac{2}{a} \: \vQ(\phi,\theta) 
\label{5.16} 
\end{equation}
at collisions. The free streaming time evolution of $\vrho$ and $\vSig$ is 
given by the differential equations 
\begin{equation}
\frac{1}{v}\frac{d\vrho}{dt} = \unit - \left(\frac{\la}{2}\right)^2 
\vrho(\vSig\vSig^\T)^{-2}\vrho, \;\;\;\;\; \frac{1}{v}\frac{d\vSig}{dt} = 
\vrho^{-1}\vSig, 
\label{5.17} 
\end{equation}
which are the three dimensional version of Eqs.~(\ref{2.24}). Since 
$\vSig^+ = \vSig^-$, the second equation in Eqs.~(\ref{5.17}) can be 
integrated to get 
\begin{equation}
\vSig_t = \mathcal{T}\exp\left( v \int_0^t d\tau \vrho_\tau^{-1} 
\right)\vSig_0, 
\label{5.18} 
\end{equation}
where $\mathcal{T}$ stands for the time ordering operator. Finally, taking 
the determinant of both sides of Eq.~(\ref{5.18}) we recover 
Eq.~(\ref{5.12}), namely $\sigma_t^2 = \sigma_0^2 \,\exp(t h_t)$ with 
\begin{equation} 
h_t = \frac{v}{t} \int_0^t d\tau \: \mathrm{tr}\vrho_\tau^{-1}. 
\label{5.19} 
\end{equation}
 
We can also calculate the wave packet auto-correlation function, $C(t)$, 
defined in Eq.~(\ref{1.1}), on periodic orbits for times $t$ given by 
Eq.~(\ref{3.1}). The algebra is straightforward, but rather lengthy, and we 
provide here only the main result of the calculation: the auto-correlation 
function, $C(t)$, exhibits a sequence of sharp maxima, periodic orbit 
revivals, which occur at times $t_\mathrm{max}=nT$, with $C_\mathrm{max}
\sim\sigma_{t_\mathrm{max}}^{-2}$, so that the envelope of the PO revivals 
shows mainly exponential decay with the rate given by the KS-entropy, 
$C_\mathrm{max} \sim \exp(-h_\mathrm{KS} t_\mathrm{max})$.

\section{Summary} 
 
In this paper we have considered the short time spreading of a small 
Gaussian wave packet for a particle moving in an array of fixed, hard 
sphere scatterers, in both two and three dimensions. Our calculations are 
based upon the semi-classical expression for the quantum propagator in 
terms of the classical action for paths of the particle. We find that for 
times less than the Ehrenfest time, the spreading of the quantum wave 
packet is determined by the sum of the positive Lyapunov exponents that 
describe the classical separation of nearby trajectories. We used the 
expressions for the propagator to calculate the wave packet 
auto-correlation function for periodic orbits. Our results agree with 
earlier results of Heller \cite{heller}: (1) this function exhibits a set 
of sharp maxima, the periodic orbit revivals, whenever the moving wave 
packet overlaps with the initial one and has the same velocity direction; 
and (2) The strengths of the maxima decrease exponentially with a decay 
rate given by the positive Lyapunov exponents. When the velocities are 
oppositely directed, the correlation function takes on extremely small 
values, even though the wave packets spatially overlap. Finally we used a 
special property of the eigenfunctions for hard sphere Lorentz gases to 
evaluate the quantum fidelity, or Loschmidt echo, for a perturbing 
Hamiltonian that is just a small change in the mass of the moving particle. 
The property that the eigenfunctions are independent of the mass of the 
particle, when expressed in terms of the wave number, allowed us to relate 
the Loschmidt echo at long times to the wave packet auto-correlation 
function at much shorter times. Therefore, for periodic orbits, at least, 
the Loschmidt echo will exhibit the same kind of periodic orbit revivals as 
one finds for the correlation functions. 
 
It would be very interesting if one could provide analytic calculations of 
quantum echos and revivals over longer time intervals for Lorentz gases or 
other, simpler models, such as quantum multibaker models \cite{wd}.  We 
would need to find appropriate techniques for analyzing the space and time 
development of wave packets for times longer than the Ehrenfest time. This 
would enable us to describe the numerical results for the Lyapunov decay 
obtained by Pastawski, Jalabert and co-workers \cite{past,cucc}.  It is not 
yet clear to us how this might be accomplished.

\vspace{2.5cm} 
{\bf{ACKNOWLEDGMENT}} 
The authors would like to thank Daniel Wojcik, Henk van Beijeren, Pierre 
Gaspard and Ilya Arakelyan for helpful conversations. JRD wishes to thank
the National Science Foundation for support under grant PHY-01-38697.

\end{document}